\documentclass[aps, prd, preprint,tightenlines, showpacs,nofootinbib,superscriptaddress]{revtex4}

\usepackage{amssymb}
\usepackage{amsmath}
\usepackage{epsfig}

\begin{document}

\title{BFKL and Sudakov Resummation in Higgs Boson Plus Jet Production with Large
Rapidity Separation}

\author{Bo-Wen Xiao}
\affiliation{Key Laboratory of Quark and Lepton Physics (MOE) and Institute
of Particle Physics, Central China Normal University, Wuhan 430079, China}

\author{Feng Yuan}
\affiliation{Nuclear Science Division, Lawrence Berkeley National
Laboratory, Berkeley, CA 94720, USA}

\begin{abstract}
We investigate the QCD resummation for the Higgs boson plus a high $P_T$ jet
production with large rapidity separations in proton-proton collisions at the LHC. The relevant 
Balitsky-Fadin-Kuraev-Lipatov (BFKL) and Sudakov logs are identified 
and resummed. In particular, we apply recent developments of the 
transverse momentum dependent factorization formalism
in the impact factors, which provides a systematic framework to 
incorporate both the BFKL and Sudakov resummations.   
\end{abstract}
\maketitle

{\it Introduction.}
The production of a Higgs boson in association with a large 
transverse momentum jet is an important channel
at the LHC to investigate the Higgs boson property,
in particular, when they are produced with large rapidity 
separation~\cite{Aad:2012tfa,Chatrchyan:2012ufa,Dittmaier:2012vm}. 
To explore the full potential to distinguish
between different production mechanisms, we need to 
improve the theoretical computations of this process. There
have been great progresses in higher order perturbative 
calculations in the last few years with next-to-next-to-leading
order results available~\cite{Chen:2014gva,Boughezal:2015dra,Boughezal:2015aha,Caola:2015wna,Chen:2016zka}. 
In addition, there exist large logarithms to be resummed to all orders
to make reliable theoretical predictions. Because of the large rapidity separation
between the two final state particles, an important contribution
comes from the so-called Balitsky-Fadin-Kuraev-Lipatov 
(BFKL) evolution~\cite{Balitsky:1978ic}, similar to the Mueller-Navelet (MN)
dijet production~\cite{Mueller:1986ey}. Meanwhile, there are
Sudakov-type of large logarithms~\cite{Sudakov:1954sw,Collins:1984kg}, 
which has been shown in Ref.~\cite{Sun:2014lna} for central rapidity 
Higgs boson plus jet production. In this paper, we will 
develop a systematic framework to implement both BFKL and
Sudakov resummations for the Higgs boson plus jet 
production with large rapidity separation at the LHC. 
This is crucial for phenomenological study to 
investigate the coupling between the 
Higgs boson and  other particles in the Standard Model.

We focus on the QCD process contributions to the Higgs boson
plus jet production~\footnote{For the electroweak process, such as
vector boson fusion contributions, there is no BFKL type of
logarithms at higher orders, though the QCD-Sudakov double logarithms 
still exist~\cite{Sun:2016mas}.},
\begin{equation}
p(P_A)+p(P_B)\to H(y_1,k_{1\perp})+Jet(y_2,k_{2\perp}) \ , \label{process}
\end{equation}
where the incoming hadrons carry momenta $P_A$ and $P_B$, 
two final state particles with rapidities $y_1$ and $y_2$, 
transverse momenta $k_{1\perp}$ and $k_{2\perp}$,
respectively. We take the limit of large rapidity difference 
$Y=|y_1-y_2| \sim \frac{1}{\alpha_s}\gg 1$ , where as schematically shown in Fig.~1, 
we can write down the following factorization formula in the momentum space as follows
\begin{eqnarray}
\frac{d^6\sigma(pp\to H+J)}{dy_1dy_2d^2k_{1\perp}d^2k_{2\perp}}=\sum_{b=q,g}
\int d^2q_{1\perp}d^2q_{2\perp}V_h(x_1,q_{1\perp},k_{1\perp})
V_b(x_2,q_{2\perp},k_{2\perp})f_{BFKL}(q_{1\perp},q_{2\perp};Y) \ , \label{hjet}
\end{eqnarray}
where $V_b$ is the impact factor for parton $b$ (quark or gluon), $V_h$
for the Higgs boson, and $f_{BFKL}$ represents the BFKL
evolution effects due to gluon radiation in the rapidity interval of $Y$.
This factorization is very much similar to the MN-dijet production 
process~\cite{Mueller:1986ey}, where the dijet are well separated in 
rapidity. There have been great progresses in theory developments for MN-dijet
productions~\cite{Fadin:1998py,Ciafaloni:1998kx,Ciafaloni:1998hu,Bartels:2001ge,Bartels:2002yj,
Colferai:2010wu,Caporale:2011cc,Ducloue:2013hia,Ducloue:2013bva,Caporale:2014gpa},
and the first detailed experiment measurement have been performed by
the CMS collaboration at the LHC~\cite{Khachatryan:2016udy}. 
The experimental results have been interpreted as an evidence
for the BFKL dynamics~\cite{Ducloue:2013bva}. In our previous
publication, we have shown that there exist Sudakov logarithms
in MN-dijet productions and these logarithms should be resummed as well~\cite{Mueller:2015ael}.
Our results in the following can be applied to MN-dijet processes,
and will confirm the factorization formula postulated there. The important difference between the Higgs+Jet process 
and the MN dijet process is that the Higgs mass can serve as an additional scale which makes the Sudakov resummation a 
bit more non-trivial. Using the Fourier transform, it is straightforward to write the above factorization formula in the coordinate space, where the 
resummation is performed
\begin{eqnarray}
\frac{d^6\sigma(pp\to H+J)}{dy_1dy_2d^2k_{1\perp}d^2k_{2\perp}}&=&\sum_b
\int \frac{d^2b_{1\perp}d^2b_{2\perp}}{(2\pi)^4} e^{ik_{1\perp}\cdot b_{1\perp}+i k_{2\perp}\cdot b_{2\perp}} \notag \\
&&\times \widetilde{V}_h(x_1,b_{1\perp},k_{1\perp})
\widetilde{V}_b(x_2,b_{2\perp},k_{2\perp}) \widetilde{f}_{BFKL}(b_{1\perp},b_{2\perp};Y)  \ . \label{hjetc}
\end{eqnarray}
It is well-known that both the Sudakov resummation and BFKL evolution can be more conveniently carried out in the coordinate space. 

\begin{figure}[tbp]
\begin{center}
\includegraphics[width=8.0cm]{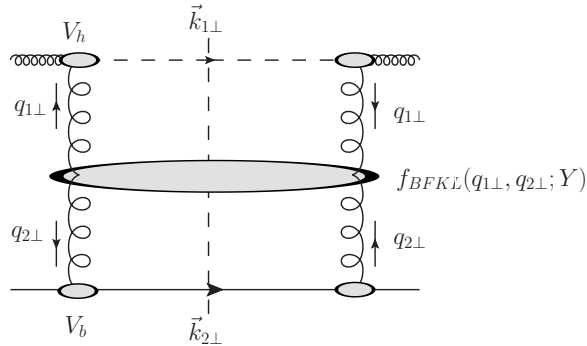}
\end{center}
\caption[*]{Schematic factorization for Higgs boson and
a hard jet production with large rapidity separation between them 
at the LHC: $f_{BFKL}$ represents the BFKL evolution with gluon
radiation in the rapidity interval between the two final state particles; 
$V_h$ and $V_b$ for the transverse momentum resummation 
effects with gluon radiation in the forward rapidity region of the
incoming gluon and partons, respectively. }
\label{factorization}
\end{figure}

In the inclusive production process, the impact factors can be
calculated in the collinear factorization approach.
However, in the study of the azimuthal angular distribution
between the two final state particles, there
exist Sudakov double logarithms in the back-to-back
correlation kinematics, where, for example, $k_{1\perp}$ 
is close to $q_{1\perp}$. To resum these large Sudakov type logarithms, we apply
the transverse momentum dependent (TMD) factorization~\cite{Collins:1984kg,Collins:2011zzd,Ji:2004wu} for the
impact factors in Eq.~(\ref{hjet}): $V_h$ is the TMD gluon
distribution, and $V_b$ is factorized into the TMD parton distribution
and the soft factor associated with the final state jet. The resummation
is carried out by solving the relevant evolution equation.

The physical argument for the above factorization is that
the higher order gluon radiations can be classified according to the relevant 
phase space. The most important
gluon radiation comes from the large rapidity separation region between the two final state particles,
which generates the BFKL evolution effects and can be factorized
into the factor $f_{BFKL}$. In the meantime,
the gluon radiations in the forward regions of the incoming quark and gluon 
are factorized into the TMD parton distributions, with a manifest
rapidity cut-off in their definitions~\cite{Collins:2011zzd}.
Therefore, the BFKL and Sudakov contributions are clearly
separated out in the gluon radiation phase space and the factorization
can be proved accordingly.
This will build a systematic framework to implement both BFKL and
Sudakov resummation in the process of Eq.~(\ref{process}). 

From the resummation point of view, there are two interesting types of logarithms arising from a one-loop
calculation for this process, namely, the BFKL type logarithm $\alpha_s Y$ and the Sudakov logarithms. 
They can be resummed into the factor $f_{BFKL}$ and the impact factors, respectively. As far as the collinear logarithms 
are concerned, they can be easily dealt with the help of the jet definition and the collinear parton distributions. 

The rest of the paper is organized as follows. We take 
the example of the quark impact factor to demonstrate how
the TMD factorization and resummation are applied.
Similar results can be obtained for the gluon impact 
factor. We then calculate the Higgs impact factor,
which is factorized into the TMD gluon distribution. Finally,
we summarize our results.

{\it Impact Factors for the Quark and Gluon.}
The partonic scattering of the process described in Eq.~(\ref{process}) comes
from quark-gluon and gluon-gluon channels. 
According to the proposed BFKL factorization, we can separate the
calculations into the quark or gluon impact factor and the gluon-Higgs
impact factor. Let us take the quark impact factor as an example. At one-loop order,
there are virtual and real gluon radiation contributions. The virtual
contribution can be written as
\begin{eqnarray}
\Gamma^v=\frac{\alpha_s}{2\pi}\left(\frac{\mu^2}{\vec{q}_\perp^2}\right)^\epsilon
\left\{C_F\left[-\frac{2}{\epsilon^2}-\frac{3}{\epsilon}-\frac{23}{4}+\frac{3}{2}\pi^2\right]+{\cal K}\right\} \ , \label{qvirtual}
\end{eqnarray}
where ${\cal K}=C_A\left(\frac{67}{18}-\frac{\pi^2}{6}\right)-\frac{5}{9}N_f$, 
$C_F=4/3$ and $N_c=3$, $N_f$ represents the number of quark flavors.
Here we work in the dimensional regulation with $D=4-2\epsilon$ and $\overline{\textrm{MS}}$ scheme. 
In the above equation, $\vec{q}_\perp$ is the t-channel momentum transfer due to the BFKL factor, and 
a universal energy dependent term proportional to
$C_A\ln(s_0/\vec{q}_\perp^2)/\epsilon$ is omitted\footnote{It is very clear that this term corresponds to the BFKL dynamics, since it is proportional to $C_A$ instead of $C_F$ and it depends on the collision energy. It is well-known that the BFKL evolution equation is an energy evolution equation which is proportional to $C_A$.}. Together with the similar
term from the real gluon radiation, it generates the corresponding BFKL 
contribution, which can be used to derive the well-known BFKL evolution equation. The detailed procedure can be found in Ref.~\cite{Mueller:2015ael, Watanabe:2016gws}. In the following, we will
focus on the QCD dynamics associated with the Sudakov logarithms, and
neglect the BFKL part to simplify the derivations. 
The real gluon radiation amplitude has also been calculated,
\begin{eqnarray}
\frac{\alpha_s}{2\pi^2}
\int d^2k_\perp \left[\frac{1+z^2}{1-z}-\epsilon (1-z)\right]\left\{C_F\frac{(1-z)^2q_\perp^2}{k_\perp^2(k_\perp-(1-z)q_\perp)^2}\right\} \ ,
\end{eqnarray}
where $(1-z)$ is the momentum fraction of the incoming quark carried by the
radiated gluon with transverse momentum $k_\perp$. Clearly, there are
two important contributions from the singularities in the above equation:
(1) collinear gluon radiation associated with the incoming quark when $k_\perp\to 0$;
(2) soft gluon radiation associated with the final state jet when $k_\perp\sim (1-z) q_\perp$.
We take the leading power contribution in the limit of $k_\perp\ll q_\perp$,
where soft gluon radiation with $z\to 1$ plays an important role.
By apply the plus function prescription to separate out the
collinear gluon radiation from the incoming quark, we are left 
with the following term,
\begin{equation}
\frac{\alpha_s}{2\pi^2}C_F\frac{1}{k_\perp^2}
\delta(1-z)
\int\frac{d\alpha}{\alpha}(1+(1-\alpha)^2)\frac{\alpha^2 q_\perp^2}{(k_\perp-\alpha q_\perp)^2}\ ,
\end{equation}
which contains the collinear divergence associated with the final state jet.
Following the same procedure described in Refs.~\cite{Mueller:2013wwa,Sun:2014gfa,Sun:2015doa}, we apply the anti-$k_t$
jet algorithm and the narrow jet approximation~\cite{Jager:2004jh,Mukherjee:2012uz} which lead to
\begin{equation}
\frac{\alpha_s}{2\pi^2}\frac{1}{k_\perp^2}\left[\ln\frac{q_\perp^2}{k_\perp^2}+\ln\frac{1}{R^2}
+\epsilon \left(\frac{1}{2}\ln^2\frac{1}{R^2}+\frac{\pi^2}{6}\right)\right] \ ,
\end{equation}
where $R$ represents the jet size. 
The soft divergence in the above equation will be cancelled out by
the virtual contribution in Eq.~(\ref{qvirtual}). To see this more clearly, 
we introduce the Fourier transform in $b_\perp$-space: $V_q(x,k_{1\perp},q_{1\perp})=\int\frac{d^2b_\perp}{(2\pi)^2}e^{i(k_{1\perp}-q_{1\perp})\cdot b_\perp}\widetilde{V}_q(x,b_\perp)$, and write the one-loop result for $\widetilde{V}_q$ as
\begin{eqnarray}
\widetilde{V}_q^{(1)}(b_\perp)&=&\widetilde{V}_q^{(0)}(b_\perp)
\frac{\alpha_s}{2\pi}\left\{C_F{\cal P}_{qq}(z)\left(-\frac{1}{\epsilon}-\ln\frac{q_\perp^2b_\perp^2}{c_0^2}\right)-(1-z)C_F\right.\\
&&\left.+\delta(1-z)\left[C_F\left(-\frac{1}{2}\ln^2\left(\frac{q_\perp^2b_\perp^2}{c_0^2}\right)
+\left(\frac{3}{2}-\ln\frac{1}{R^2}\right)\ln\frac{q_\perp^2b_\perp^2}{c_0^2}\right)+{\cal K}+\Delta I_q\right]\right\}\ ,
\end{eqnarray}
where $\widetilde{V}_q^{(0)}$ represents the leading order normalization, 
$c_0=2e^{-\gamma_E}$, ${\cal P}_{qq}(z)$ is the quark-quark splitting kernel and $\Delta I_q=C_F\left[\frac{3}{2}\ln\frac{1}{R^2}+\frac{3}{4}+\frac{2}{3}\pi^2\right]$. In reaching the above expression, we have also included the jet contribution~\cite{Sun:2015doa}. 
Clearly, there are Sudakov double and single logarithms.
The above result can be factorized into the TMD quark 
distribution and the soft factor associated with the final state
jet. Here we follow the Collins 2011 scheme for the definition of TMDs, which
are defined with soft factor subtraction~\cite{Collins:2011zzd} as follows
\begin{equation}
f_{q}^{(sub.)}(x,b_\perp,\mu_F,\zeta_c)=f_q^{unsub.}(x,b_\perp)\sqrt{\frac{S_2^{\bar
n,v}(b_\perp)}{S_2^{n,\bar n}(b_\perp)S_2^{n,v}(b_\perp)}} \ , \label{jcc}
\end{equation}
where $b_\perp$ is the Fourier conjugate variable respect to
the transverse momentum $k_\perp$, $ \mu_F$ the factorization
scale and $\zeta_c^2=x^2(2v\cdot P)^2/v^2=2(xP^+)^2e^{-2y_n}$ with
$y_n$ the rapidity cut-off in the Collins-2011 scheme. The second factor corresponds to the 
soft factor subtraction with $n$ and $\bar n$ as the light-front vectors $n=(1^-,0^+,0_\perp)$,
$\bar n=(0^-,1^+,0_\perp)$, whereas $v$ is an off-light-front four-vector $v=(v^-,v^+,0_\perp)$ with $v^-\gg v^+$. 
The un-subtracted TMD reads as 
 \begin{eqnarray}
f_q^{unsub.}(x,k_\perp)&=&\frac{1}{2}\int
        \frac{d\xi^-d^2\xi_\perp}{(2\pi)^3}e^{-ix\xi^-P^++i\vec{\xi}_\perp\cdot
        \vec{k}_\perp}  \left\langle
PS\left|\overline\psi(\xi){\cal L}_{n}^\dagger(\xi)\gamma^+{\cal L}_{n}(0)
        \psi(0)\right|PS\right\rangle\ ,\label{tmdun}
\end{eqnarray}
with the gauge link defined as $ {\cal L}_{n}(\xi) \equiv \exp\left(-ig\int^{-\infty}_0 d\lambda
\, v\cdot A(\lambda n +\xi)\right)$. The light-cone singularity in the un-subtracted
TMDs is cancelled out by the soft factor as in Eq.~(\ref{jcc}) with $S^{v_1,v_2}$
defined as
\begin{equation}
S_2^{v_1,v_2}(b_\perp)={\langle 0|{\cal L}_{v_2}^\dagger(b_\perp) {\cal
L}_{v_1}^\dagger(b_\perp){\cal L}_{v_1}(0){\cal
L}_{v_2}(0)  |0\rangle   }\, . \label{softg}
\end{equation}
Following the similar idea, we introduce a subtracted soft factor associated
with the final state jet,
\begin{eqnarray}
S_J(b_\perp,\mu_F)=\sqrt{\frac{S_{n,n_1}(b_\perp)S_{n_1,\bar n}(b_\perp)}{S_{n,\bar n}(b_\perp)} }\ ,
\end{eqnarray}
where $n_1$ represents the jet direction. One-loop calculation leads
to the following result,
\begin{equation}
S_J^{(1)}=\frac{\alpha_s}{2\pi}C_F\left[\ln\frac{1}{R^2}
\ln\frac{b_\perp^2\mu_F^2}{c_0^2}+\frac{1}{2}\ln^2\left(\frac{1}{R^2}\right)+\frac{\pi^2}{6}\right] \ ,
\end{equation}
again with narrow jet approximation, 
from which we obtain the anamolous dimension $\gamma^{(s)}=\frac{\alpha_s}{2\pi} C_F\ln(1/R^2)$. 
Together with the result for the quark distribution from Ref.~\cite{Collins:2011zzd,Sun:2013hua}, the following TMD
factorization can be verified at one-loop order,
\begin{equation}
\widetilde{V}_q(x,b_\perp)=
f_{q}^{(sub.)}(x,b_\perp,\mu_F,\zeta_c) S_J(b_\perp,\mu_F) H(q_{1\perp},\mu_F) \ .
\end{equation}
Furthermore, in order to eliminate the large logarithms in the hard factor $H^{(1)}$, 
we have to choose the appropriate scales as $\mu_F^2=\zeta_c^2=\vec{q}_\perp^2$.
This corresponds to the factorization that the TMD quark distribution only contains contribution 
from the gluon radiation in the forward region of the incoming quark. The gluon radiation
in the central region (rapidity interval between the two final state particles) belongs
to the BFKL evolution. Finally, following the Collins-Soper-Sterman (CSS) 
resummation approach~\cite{Collins:1984kg}, we obtain the all order result as follows
\begin{eqnarray}
\widetilde{V}_q(x,b_\perp)&=&\widetilde{V}_q^{(0)}e^{-S_q(q_\perp,b_\perp)} C\otimes 
f_q(x,\bar \mu={c_0}/{b})\left[1+\frac{\alpha_s}{2\pi}\left({\cal K}+\Delta I_q\right)\right] \ ,\label{vq}
\end{eqnarray}
where $\otimes$ represents the convolution in $x$ and $f_q(x,\bar \mu)$ the
integrated quark distribution. Following the so-called ``TMD" 
scheme~\cite{Catani:2000vq,Prokudin:2015ysa}
in CSS resummation, the hard and soft factors at the appropriate scale lead to the 
coefficients at $\alpha_s$, represented by
${\cal K}$ and $\Delta I_q$. The Sudakov factor can be written as
\begin{equation}
S_q(q_\perp,b_\perp)=\int_{c_0^2/b_\perp^2}^{q_\perp^2}
\frac{d\mu^2}{\mu^2}\left[A_q\ln\frac{q_\perp^2}{\mu^2}+B_q+D_q\ln\frac{1}{R^2}\right] \ ,
\end{equation}
with $A_q=\sum_i  A_q^{(i)}$, $A_q^{(1)}=D_q^{(1)}=\frac{\alpha_s}{2\pi}C_F$, $B_q^{(1)}=-\frac{3}{2}A_q^{(1)}$,
and the $C$ coefficient function is $C^{(1)}=\frac{\alpha_s}{2\pi}C_F(1-x)$. 

Similar calculations can be performed for the gluon impact factor,
\begin{eqnarray}
\widetilde{V}_g(x,b_\perp)&=&\widetilde{V}_g^{(0)}e^{-S_g(q_\perp,b_\perp)} C\otimes 
f_g(x,\bar \mu={c_0}/{b})\left[1+\frac{\alpha_s}{2\pi}\left({\cal K}+\Delta I_g\right)\right] \ ,\label{vg}
\end{eqnarray} 
with one-loop results as $A_g^{(1)}=D_g^{(1)}=\frac{\alpha_s}{2\pi}C_A$, $B_g^{(1)}=-2\beta_0A^{(1)}$.
and $\Delta I_g=C_A\left(2\beta_0\ln\frac{1}{R^2}-\frac{\pi^2}{6}\right)-\frac{N_f}{6}$ 
with $\beta_0=\frac{11}{12}-\frac{N_f}{18}$ and $N_f$ being the number of flavors. The $C$ coefficient
vanishes at one-loop order.
In the BFKL factorization, the quark and gluon impact factors are universal, which means that they are same as those in MN-dijet processes. 
Indeed, we can apply the above impact
factors and obtain the consistent results as those in Ref.~\cite{Mueller:2015ael}.

{\it Impact Factor for the Higgs Boson.}
The computation procedure of the last section can be applied
to the Higgs impact factor as well.
At the one-loop order, the virtual graph contribution in the gluon-to-Higgs boson impact 
factor can be deduced from that in Higgs boson 
plus jet production by taking the limit of the large rapidity separation between
the final state particles~\cite{Ravindran:2002dc,Glosser:2002gm},
\begin{eqnarray}
\Gamma^v=\frac{\alpha_s}{2\pi} 
\left(\frac{\mu^2}{\vec{q}_\perp^2}\right)^\epsilon
\left\{N_c\left[-\frac{1}{\epsilon^2}+\frac{1}{\epsilon}\left(
\ln\frac{\tilde m^2}{\vec{q}_\perp^2}
-2\beta_0\right)\right] +\Delta I_h+{\cal K}\right\} \ ,
\end{eqnarray}
where $\tilde m^2=m_h^2+\vec{q}_\perp^2$ with Higgs mass $m_h$, 
$\Delta I_h=C_A\left[\pi^2+2{\rm Li}_2(x_q)+\ln(x_q)\ln\frac{(1+x_q)^2}{x_q}\right]$
and $x_q$ is defined as $x_q=\vec{q}_\perp^2/m_h^2$. 
Again, we have subtracted the universal energy
dependent term related to the BFKL evolution. 
The contribution from the real gluon radiation can be summarized as
\begin{eqnarray}
\frac{\alpha_s}{2\pi^2}\frac{1}{k_\perp^2}C_A\left\{{\cal P}_{gg}(z)+\delta (1-z)
\left[\ln\frac{\tilde m^2}{k_\perp^2}-2\beta_0\right] \right\}\ . \label{hr}
\end{eqnarray}
Adding the above two terms together, we obtain the following result
in the $b_\perp$-space,
\begin{eqnarray}
\widetilde{V}_h^{(1)}&=&\widetilde{V}_h^{(0)}\frac{\alpha_s}{2\pi}\left[C_A{\cal P}_{gg}(z)+\delta(1-z)
\left(\Delta I_h+{\cal K}\right)\right]\notag \\
&& +\widetilde{V}_h^{(0)}\frac{\alpha_s}{2\pi} C_A \delta(1-z) \left[-\frac{1}{2}\ln^2\left(\frac{q_\perp^2b_\perp^2}{c_0^2}\right)
+\left(\ln\frac{\tilde{m}^2}{q_\perp^2}-2\beta_0\right)\ln\frac{c_0^2}{q_\perp^2b_\perp^2}\right] ,
\end{eqnarray}
where ${\cal P}_{gg}(z)$ represents the gluon-gluon splitting kernel.
Again, the above result can be factorized into the TMD gluon distribution,
\begin{equation}
\widetilde{V}_h=xf_g(x,b_\perp,\mu_F,\zeta_c)H(q_{\perp},\mu_F) \ ,
\end{equation}
for which we will choose the factorization scale $\mu_F^2=\vec{q}_\perp^2$ and $\zeta_c^2=\tilde m^2$
to eliminate the large logarithms in the hard factor. 
All order resummation is achieved by solving the energy evolution
equation for the TMD gluon distribution, 
\begin{eqnarray}
\widetilde{V}_h(x,b_\perp)&=&\widetilde{V}_h^{(0)}
e^{-S_h(q_\perp,b_\perp)} C\otimes f_g(x,\bar \mu={c_0}/{b})\left[1+\frac{\alpha_s}{2\pi}\left({\cal K}+\Delta I_h\right)\right] \ ,\label{vh}
\end{eqnarray}
where the Sudakov factor can be written as
\begin{equation}
S_h(q_\perp,b_\perp)=\int_{c_0^2/b_\perp^2}^{q_\perp^2}
\frac{d\mu^2}{\mu^2}\left[A_h\ln\frac{\tilde m^2}{\mu^2}+B_h\right] \ .  \label{sh}
\end{equation}
We find that $A_h=A_g$, $B_h=B_g$, which is because they come from the
same TMD gluon distribution, 
and the $C$ coefficient function vanishes at one-loop order.

{\it Summary and Discussions.}
The final resummation results for the BFKL and Sudakov 
resummation effects in the Higgs boson plus jet production with large 
rapidity separation are obtained by 
substituting the results in Eqs.~(\ref{vq},\ref{vg},\ref{vh}) into 
Eq.~(\ref{hjetc}). An important cross check has been performed by comparing
to the derivation in Ref.~\cite{Sun:2014lna} with only Sudakov resummation, and we find
the complete agreement. 

The factorization method developed in
this paper can have great impact in LHC physics. A potential
application is to study in Higgs plus two 
jets production where the final state three particles are well
separated in rapidity. This channel is an important place to
study the vector boson fusion contribution in Higgs boson
production at the LHC, where we need to understand the 
QCD resummation contributions accurately.

Theoretically, both BFKL and Sudakov resummations
are the important corner stones in the perturbative QCD applications
to high energy hadronic collisions. Recently, there have been
strong interests~\cite{Mueller:2013wwa, Mueller:2012uf, Balitsky:2015qba,
Marzani:2015oyb, Balitsky:2016dgz, Zhou:2016tfe, Xiao:2017yya}  
to combine these two resummations
consistently in the hard scattering processes
at various collider experiments.
Our results in this paper is a step further toward a systematic framework
to deal with both physics. We anticipate 
more applications in the future, in particular, for multi-jets events
at the LHC, such as three-jet or four-jet productions~\cite{Caporale:2016soq,Caporale:2016xku,Caporale:2016zkc}.

\section*{Acknowledgements}
We thank Al Mueller for stimulating discussions and comments. This work was supported in part by the U.S. Department of Energy under the contracts DE-AC02-05CH11231 and by the NSFC under Grant No.~11575070.

\newpage

\appendix
\section{Consistency in the BFKL Evolution Effects}

In this section, we show that the BFKL evolution has been consistently taken into 
account with the above results w.r.t. the impact factor calculations.

In the quark impact factor calculation, we have the following term associated 
with the BFKL gluon radiation,
\begin{equation}
\frac{\alpha_s}{2\pi^2}C_A\int \frac{dz}{z}\frac{2(1-z)k_\perp\cdot (k_\perp-zq_\perp)}{k_\perp^2(k_\perp-zq_\perp)^2} \ .
\end{equation}
Here $z$ integral is limited by $z>k_\perp^2/s_\Lambda$, where $s_\Lambda=2p_1^-P_A^+$ represents
the invariant mass cut-off with $p_1$ for the incoming quark momentum. 
This cut-off will be combined with the other particle in the final state
to obtain the boost invariant evolution for the BFKL gluon radiation. There is no final state jet divergence, because the
$z\sim k_\perp/q_\perp$ is regulated by the numerator. Further calculations can be performed by average
the azimuthal angle between $k_\perp$ and $q_\perp$, from which we find the integrand vanishes 
in the region of $z>k_\perp/q_\perp$. Therefore, the final result will be
\begin{equation}
\frac{\alpha_s}{2\pi^2}C_A\frac{1}{k_\perp^2}\ln\frac{s_\Lambda^2}{q_\perp^2 k_\perp^2}  \ .
\end{equation}
In the case of Mueller-Navelet dijet productions, we can perform the same calculation 
for the other impact factor and introduce $\bar s_\Lambda=2p_2^+P_B^-$ with the following expression,
\begin{equation}
\frac{\alpha_s}{2\pi^2}C_A\frac{1}{k_\perp^2}\ln\frac{\bar s_\Lambda^2}{q_\perp^2 k_\perp^2}  \ .
\end{equation}
By taking into account the kinematic relation $s_\Lambda\bar s_\Lambda=ss_y=s_y^2$ where 
$s_y=q_\perp^2e^{\Delta Y}$, we can find the BFKL evolution simplifies to
\begin{equation}
\frac{\alpha_s}{2\pi^2}C_A\frac{2}{k_\perp^2}\ln\frac{s_y^2}{q_\perp^2 k_\perp^2}  \ .
\end{equation}
The above is the universal BFKL evolution contribution, which only depends on the transverse
momentum and the rapidity between the two final state particles. We expect the
same BFKL contribution from the Higgs plus jet production as well. This provides
an important cross check for the above calculations.

From the details of the gluon-Higgs impact factor calculation, we find that there is only the
following term contributing to the BFKL evolution,
\begin{equation}
\frac{\alpha_s}{2\pi^2}C_A\frac{2}{k_\perp^2}\ln\frac{s_\Lambda}{k_\perp^2}\ ,
\end{equation}
and all other power suppressed terms drop out from the calculations in Ref.~\cite{Sun:2014lna}.
It is interesting to note that the above term can be separated into two terms,
\begin{equation}
\frac{\alpha_s}{2\pi^2}C_A\frac{2}{k_\perp^2}\ln\frac{s_\Lambda}{k_\perp^2} = 
\frac{\alpha_s}{2\pi^2}C_A\frac{1}{k_\perp^2}\left[\ln\frac{\tilde m^2}{k_\perp^2} +\ln\frac{s_\Lambda^2}{\tilde m^2 k_\perp^2}\right]\ ,
\end{equation}
where the first term contributes to the Sudakov logs (as in Eqs.~(\ref{hr}) and (\ref{sh})), and the second term gives the BFKL evolution after combined with
the BFKL term from the quark impact factor calculation. The latter is achieved by
taking into account the following identity from the kinematics of Higgs boson plus jet production,
\begin{equation}
s_\Lambda^2 \bar s_\Lambda^2=s^2s_y^2=s_y^4\tilde m^2/q_\perp^2 \ ,
\end{equation}
in the limit of large rapidity separation between the Higgs boson and the produced jet.


\begin{thebibliography}{99}

\bibitem{Aad:2012tfa}
  G.~Aad {\it et al.}  [ATLAS Collaboration],
  Phys.\ Lett.\ B {\bf 716}, 1 (2012).
  
  
\bibitem{Chatrchyan:2012ufa}
  S.~Chatrchyan {\it et al.}  [CMS Collaboration],
  Phys.\ Lett.\ B {\bf 716}, 30 (2012).
 

\bibitem{Dittmaier:2012vm}
  S.~Dittmaier, S.~Dittmaier, C.~Mariotti, G.~Passarino, R.~Tanaka, S.~Alekhin, J.~Alwall and E.~A.~Bagnaschi {\it et al.},
  arXiv:1201.3084 [hep-ph];
  
  S.~Heinemeyer {\it et al.}  [LHC Higgs Cross Section Working Group Collaboration],
  arXiv:1307.1347 [hep-ph].
\bibitem{Chen:2014gva} 
  X.~Chen, T.~Gehrmann, E.~W.~N.~Glover and M.~Jaquier,
  Phys.\ Lett.\ B {\bf 740}, 147 (2015)
  doi:10.1016/j.physletb.2014.11.021
  [arXiv:1408.5325 [hep-ph]].

\bibitem{Boughezal:2015dra}
  R.~Boughezal, F.~Caola, K.~Melnikov, F.~Petriello and M.~Schulze,
  Phys.\ Rev.\ Lett.\  {\bf 115}, no. 8, 082003 (2015).
\bibitem{Boughezal:2015aha}
  R.~Boughezal, C.~Focke, W.~Giele, X.~Liu and F.~Petriello,
  Phys.\ Lett.\ B {\bf 748}, 5 (2015)
  
\bibitem{Caola:2015wna}
  F.~Caola, K.~Melnikov and M.~Schulze,
  Phys.\ Rev.\ D {\bf 92}, no. 7, 074032 (2015)
  
\bibitem{Chen:2016zka} 
  X.~Chen, J.~Cruz-Martinez, T.~Gehrmann, E.~W.~N.~Glover and M.~Jaquier,
  JHEP {\bf 1610}, 066 (2016)
  doi:10.1007/JHEP10(2016)066
  [arXiv:1607.08817 [hep-ph]].
  
\bibitem{Balitsky:1978ic} 
  I.~I.~Balitsky and L.~N.~Lipatov,
  Sov.\ J.\ Nucl.\ Phys.\  {\bf 28}, 822 (1978)
  [Yad.\ Fiz.\  {\bf 28}, 1597 (1978)];
  E.~A.~Kuraev, L.~N.~Lipatov and V.~S.~Fadin,
  Sov.\ Phys.\ JETP {\bf 45}, 199 (1977)
  [Zh.\ Eksp.\ Teor.\ Fiz.\  {\bf 72}, 377 (1977)].


\bibitem{Mueller:1986ey} 
  A.~H.~Mueller and H.~Navelet,
  Nucl.\ Phys.\ B {\bf 282}, 727 (1987).
  

\bibitem{Sudakov:1954sw} 
  V.~V.~Sudakov,
  Sov.\ Phys.\ JETP {\bf 3}, 65 (1956)
  [Zh.\ Eksp.\ Teor.\ Fiz.\  {\bf 30}, 87 (1956)];
  Y.~L.~Dokshitzer, D.~Diakonov and S.~I.~Troian,
  Phys.\ Rept.\  {\bf 58}, 269 (1980);
  G.~Parisi and R.~Petronzio,
  Nucl.\ Phys.\ B {\bf 154}, 427 (1979).
\bibitem{Collins:1984kg} 
  J.~C.~Collins, D.~E.~Soper and G.~F.~Sterman,
  Nucl.\ Phys.\ B {\bf 250}, 199 (1985).
  
\bibitem{Sun:2014lna}
  P.~Sun, C.-P.~Yuan and F.~Yuan,
  Phys.\ Rev.\ Lett.\  {\bf 114}, no. 20, 202001 (2015).
 
 
\bibitem{Sun:2016mas} 
  P.~Sun, C.-P.~Yuan and F.~Yuan,
  Phys.\ Lett.\ B {\bf 762}, 47 (2016)
  doi:10.1016/j.physletb.2016.09.005
  [arXiv:1605.00063 [hep-ph]].
    
    
    \bibitem{Fadin:1998py} 
  V.~S.~Fadin and L.~N.~Lipatov,
  Phys.\ Lett.\ B {\bf 429}, 127 (1998)
  [hep-ph/9802290].


  
  %
\bibitem{Ciafaloni:1998kx} 
  M.~Ciafaloni,
  Phys.\ Lett.\ B {\bf 429}, 363 (1998)
  [hep-ph/9801322].
  
\bibitem{Ciafaloni:1998hu} 
  M.~Ciafaloni and D.~Colferai,
  Nucl.\ Phys.\ B {\bf 538}, 187 (1999)
  [hep-ph/9806350].
  
\bibitem{Bartels:2001ge} 
  J.~Bartels, D.~Colferai and G.~P.~Vacca,
  Eur.\ Phys.\ J.\ C {\bf 24}, 83 (2002)
  [hep-ph/0112283].
  
\bibitem{Bartels:2002yj} 
  J.~Bartels, D.~Colferai and G.~P.~Vacca,
  Eur.\ Phys.\ J.\ C {\bf 29}, 235 (2003)
  [hep-ph/0206290].
 
\bibitem{Colferai:2010wu} 
  D.~Colferai, F.~Schwennsen, L.~Szymanowski and S.~Wallon,
  JHEP {\bf 1012}, 026 (2010)
  [arXiv:1002.1365 [hep-ph]].
  
\bibitem{Caporale:2011cc} 
  F.~Caporale, D.~Y.~Ivanov, B.~Murdaca, A.~Papa and A.~Perri,
  JHEP {\bf 1202}, 101 (2012)
  [arXiv:1112.3752 [hep-ph]].
  
   
  %
\bibitem{Ducloue:2013hia} 
  B.~Ducloue, L.~Szymanowski and S.~Wallon,
  JHEP {\bf 1305}, 096 (2013)
  [arXiv:1302.7012 [hep-ph]].
  
   \bibitem{Ducloue:2013bva} 
  B.~Ducloue, L.~Szymanowski and S.~Wallon,
  Phys.\ Rev.\ Lett.\  {\bf 112}, 082003 (2014)
  [arXiv:1309.3229 [hep-ph]].

  
\bibitem{Caporale:2014gpa} 
  F.~Caporale, D.~Y.~Ivanov, B.~Murdaca and A.~Papa,
  Eur.\ Phys.\ J.\ C {\bf 74}, 3084 (2014)
  [arXiv:1407.8431 [hep-ph]].
  
\bibitem{Khachatryan:2016udy} 
  V.~Khachatryan {\it et al.} [CMS Collaboration],
  arXiv:1601.06713 [hep-ex].
  
  
\bibitem{Mueller:2015ael} 
  A.~H.~Mueller, L.~Szymanowski, S.~Wallon, B.~W.~Xiao and F.~Yuan,
  JHEP {\bf 1603}, 096 (2016)
  [arXiv:1512.07127 [hep-ph]].  
  
  
  
  
\bibitem{Collins:2011zzd} 
  J.~Collins,
   Foundations of Perturbative QCD, Cambridge University Press, Cambridge U.K. (2011).
  
  
  
\bibitem{Ji:2004wu} 
  X.~Ji, J.~P.~Ma and F.~Yuan,
  Phys.\ Rev.\ D {\bf 71}, 034005 (2005)
  doi:10.1103/PhysRevD.71.034005
  [hep-ph/0404183].
  
\bibitem{Watanabe:2016gws} 
  K.~Watanabe and B.~W.~Xiao,
  Phys.\ Rev.\ D {\bf 94}, no. 9, 094046 (2016)
  [arXiv:1607.04726 [hep-ph]].

  
\bibitem{Mueller:2013wwa} 
  A.~H.~Mueller, B.~W.~Xiao and F.~Yuan,
  Phys.\ Rev.\ D {\bf 88}, no. 11, 114010 (2013)
  doi:10.1103/PhysRevD.88.114010
  [arXiv:1308.2993 [hep-ph]].
  
\bibitem{Sun:2014gfa} 
  P.~Sun, C.-P.~Yuan and F.~Yuan,
  Phys.\ Rev.\ Lett.\  {\bf 113}, no. 23, 232001 (2014)
  doi:10.1103/PhysRevLett.113.232001
  [arXiv:1405.1105 [hep-ph]].
  
\bibitem{Sun:2015doa} 
  P.~Sun, C.-P.~Yuan and F.~Yuan,
  Phys.\ Rev.\ D {\bf 92}, no. 9, 094007 (2015)
  doi:10.1103/PhysRevD.92.094007
  [arXiv:1506.06170 [hep-ph]].
  
\bibitem{Jager:2004jh}
  B.~Jager, M.~Stratmann and W.~Vogelsang,
  Phys.\ Rev.\ D {\bf 70}, 034010 (2004).

\bibitem{Mukherjee:2012uz}
  A.~Mukherjee and W.~Vogelsang,
  Phys.\ Rev.\ D {\bf 86}, 094009 (2012).
 
  

  

\bibitem{Sun:2013hua} 
  P.~Sun and F.~Yuan,
  Phys.\ Rev.\ D {\bf 88}, no. 11, 114012 (2013)
  doi:10.1103/PhysRevD.88.114012
  [arXiv:1308.5003 [hep-ph]].
  
\bibitem{Catani:2000vq} 
  S.~Catani, D.~de Florian and M.~Grazzini,
  Nucl.\ Phys.\ B {\bf 596}, 299 (2001)
  [hep-ph/0008184];
  S.~Catani, L.~Cieri, D.~de Florian, G.~Ferrera and M.~Grazzini,
  Nucl.\ Phys.\ B {\bf 881}, 414 (2014)
  [arXiv:1311.1654 [hep-ph]].
   
  
\bibitem{Prokudin:2015ysa} 
  A.~Prokudin, P.~Sun and F.~Yuan,
  Phys.\ Lett.\ B {\bf 750}, 533 (2015)
  doi:10.1016/j.physletb.2015.09.064
  [arXiv:1505.05588 [hep-ph]].
    
\bibitem{Ravindran:2002dc}
  V.~Ravindran, J.~Smith and W.~L.~Van Neerven,
  Nucl.\ Phys.\ B {\bf 634}, 247 (2002).
  

\bibitem{Glosser:2002gm}
  C.~J.~Glosser and C.~R.~Schmidt,
  JHEP {\bf 0212}, 016 (2002).
 

  
  \bibitem{Mueller:2012uf} 
  A.~H.~Mueller, B.~W.~Xiao and F.~Yuan,
  Phys.\ Rev.\ Lett.\  {\bf 110}, no. 8, 082301 (2013)
  [arXiv:1210.5792 [hep-ph]].

\bibitem{Balitsky:2015qba} 
  I.~Balitsky and A.~Tarasov,
  JHEP {\bf 1510}, 017 (2015)
  [arXiv:1505.02151 [hep-ph]].
 
  
\bibitem{Marzani:2015oyb} 
  S.~Marzani,
  Phys.\ Rev.\ D {\bf 93}, no. 5, 054047 (2016)
  [arXiv:1511.06039 [hep-ph]].
    
\bibitem{Balitsky:2016dgz} 
  I.~Balitsky and A.~Tarasov,
  JHEP {\bf 1606}, 164 (2016)
  [arXiv:1603.06548 [hep-ph]].
 
  

\bibitem{Zhou:2016tfe} 
  J.~Zhou,
  JHEP {\bf 1606}, 151 (2016)
  [arXiv:1603.07426 [hep-ph]].
  
\bibitem{Xiao:2017yya} 
  B.~W.~Xiao, F.~Yuan and J.~Zhou,
  Nucl.\ Phys.\ B {\bf 921}, 104 (2017)
  [arXiv:1703.06163 [hep-ph]].
  

  
\bibitem{Caporale:2016soq} 
  F.~Caporale, F.~G.~Celiberto, G.~Chachamis, D.~Gordo Gómez and A.~Sabio Vera,
  Nucl.\ Phys.\ B {\bf 910}, 374 (2016)
  [arXiv:1603.07785 [hep-ph]].
  
\bibitem{Caporale:2016xku} 
  F.~Caporale, F.~G.~Celiberto, G.~Chachamis, D.~Gordo Gómez and A.~Sabio Vera,
  Eur.\ Phys.\ J.\ C {\bf 77}, no. 1, 5 (2017)
  [arXiv:1606.00574 [hep-ph]].
  
\bibitem{Caporale:2016zkc} 
  F.~Caporale, F.~G.~Celiberto, G.~Chachamis, D.~G.~Gomez and A.~Sabio Vera,
  Phys.\ Rev.\ D {\bf 95}, no. 7, 074007 (2017)
  [arXiv:1612.05428 [hep-ph]].

\end{thebibliography}
\end{document}